\documentclass[ps,prd,preprint,
superscriptaddress,preprintnumbers,eqsecnum,
showpacs,nofootinbib,nobibnotes]{revtex4}
\usepackage{amsfonts,bm}
\usepackage{graphicx}
\usepackage{pstricks}
\usepackage{color}

\newcommand{\be}{\begin{equation}}
\newcommand{\bea}{\begin{eqnarray}}
\newcommand{\ee}{\end{equation}}
\newcommand{\eea}{\end{eqnarray}}

\def\s#1{{\scriptscriptstyle #1}}

\def\eq#1{Eq.~(\ref{#1})}
\def\2eqs#1#2{Eqs.~(\ref{#1}) and~(\ref{#2})}
\def\3eqs#1#2#3{Eqs.~(\ref{#1}),~(\ref{#2}) and~(\ref{#3})}
\def\noeq#1{~(\ref{#1})}

\def\diff{\mathrm{d}}

\def\fg{\mathrm{I}\!\Gamma}
\def\G{\Gamma}

\def\gammaE{\gamma_\s{E}}

\begin{document}

\title{Canonical Transformations\\ and
Renormalization Group Invariance \\
in the presence of Non-trivial Backgrounds}
\date{January, 2012}

\author{D. Binosi}
\email{binosi@ectstar.eu}
\affiliation{European Centre for Theoretical Studies in Nuclear
  Physics and Related Areas (ECT*) and Fondazione Bruno Kessler, \\ Villa Tambosi, Strada delle
  Tabarelle 286, 
 I-38123 Villazzano (TN)  Italy}
 
\author{A. Quadri}
\email{andrea.quadri@mi.infn.it}
\affiliation{Dip. di Fisica, Universit\`a degli Studi di Milano via Celoria 16, I-20133 Milano, Italy\\
and INFN, Sezione di Milano, via Celoria 16, I-20133 Milano, Italy}

\begin{abstract}
\noindent
We show that for a SU(N) Yang-Mills theory the classical
background-quantum splitting is non-trivially deformed
at the quantum level by a canonical transformation
with respect to the Batalin-Vilkovisky bracket associated
with the Slavnov-Taylor identity of the theory.
This canonical transformation acts on all the fields (including the ghosts) and antifields; it uniquely fixes
the dependence  on the background field of all the one-particle irreducible Green's functions of the theory at hand. The approach is valid both at the perturbative and non-perturbative level, being based solely on symmetry requirements. As a practical application, we derive
the renormalization group equation in the presence
of a generic background and apply it
in the case of a SU(2) instanton. Finally, 
we explicitly calculate the one-loop deformation of the
background-quantum splitting in lowest order in the
instanton background.

\end{abstract}


\maketitle

\section{Introduction}

Though the background field method (BFM)
 has a long history in quantum field theory~\cite{DeWitt:1967ub}, its potential was fully appreciated only during the eighties when it was realized  
that in gauge theories one can
equivalently compute gauge-invariant physical quantities (like physical
$S$-matrix elements or the correlators of gauge-invariant operators)
by the ordinary Gell-Mann and Low's formula at zero background or by reconstructing physical connected amplitudes from background dependent one particle irreducible (1-PI) Green's functions via the Legendre transform with respect to (w.r.t.) the background field~\cite{Abbott:1980hw}.

The big advantage of employing the BFM over  conventional gauge-fixing schemes,  is then manifest, for in the presence of a background gauge field one can choose a (background) gauge-fixing  condition that preserves the background gauge invariance at the quantum level.
One thus obtains an additional background Ward identity for the vertex functional, yielding linear relations among 1-PI amplitudes
(unlike the Slavnov-Taylor (ST) identity, which gives more complicated bilinear
relations among the 1-PI Green's functions).
In general, the presence of this Ward identity simplifies enormously the calculations, and has been successfully 
exploited in many applications,  ranging from 
perturbative calculations in Yang-Mills theories~\cite{Abbott:1980hw,Ichinose:1981uw} and in the Standard Model~\cite{Denner:1994xt,hep-ph/0102005} to gravity and supergravity calculations~\cite{Gates:1983nr}\footnote{Notice, however,  that the background Ward identity
is no substitute to the ST identity: physical unitarity
stems from the validity of the latter identity and does not follow from the former identity alone 
\cite{Ferrari:2000yp}.}.

On the non-perturbative side, the discovery of topologically non-trivial gauge field configurations~\cite{Belavin:1975fg,'tHooft:1976fv} triggered the study of Yang-Mills theory around  non-vanishing vacua.
For instance, the confinement problem can be explained in the dual superconductor picture by condensing chromomagnetic monopoles, leading to a confinement potential via the formation of flux tubes for the chromoelectric field~\cite{dual_superconductor}. 
In the vortex condensation model  instead~\cite{vortex_condensation}, closed chromomagnetic center vortices condense and give rise to an area law for the Wilson loop (and eventually a dynamical mass for the gluon field).
More recently, the synthesis of the BFM with the pinch technique~\cite{Cornwall:1981zr,Binosi:2002ft}, has provided a non perturbative setting in which a new set of Schwinger-Dyson (SD) equations has been formulated~\cite{Binosi:2007pi} and the corresponding solutions used~\cite{Aguilar:2008xm} to study the properties of the infrared sector of Yang-Mills theories.

In a very recent paper~\cite{Binosi:2011ar} 
it has been shown that
the ST identity in the presence of a background field ({\it extended} ST identity in what follows) provides
a remarkable set of constraints on the vertex
functional $\G$ of Yang-Mills theories in the presence
of a non-trivial background $\widehat A^a_\mu$.
Since these constraints are a consequence of the ST identity,
they should be fulfilled by any implementation of the BFM, {\it e.g.}, when/if formulated on a lattice.

The main result of~\cite{Binosi:2011ar} is that the classical background-quantum
splitting is deformed at the quantum level in a 
non-trivial fashion, with the deformation controlled
by a particular 1-PI correlator, involving the covariant
derivatives of the ghost and the antighost fields.
More precisely, let us denote by $A^{*a}_\mu$
the antifield associated with the gauge field $A^a_\mu$
and by $\Omega^a_\mu$ the external ghost
source \cite{Grassi:1995wr,Ferrari:2000yp,Becchi:1999ir} that forms the BRST partner of the
background gauge field $\widehat A^a_\mu$.
At the classical level $A^{*a}_\mu$ is coupled
to the BRST variation of $A^a_\mu$, 
{\it i.e.}, the covariant derivative of the ghost field,
while $\Omega^a_\mu$ is coupled to the
covariant derivative of the antighost.

Then, in the full quantum theory
the deformed quantum-background splitting
amounts to 
a background-dependent
field redefinition~\cite{Binosi:2011ar}  
\bea
A^a_\mu \rightarrow A^a_\mu - {\cal G}^a_\mu(\widehat A) 
\label{bkg.split}
\eea
where the functional ${\cal G}^a_\mu$ is obtained from
the correlator $\G_{\Omega^a_\mu A^{*b}_\nu}$ via the defining
equation
\bea
\frac{\delta {\cal G}^b_\nu(y)}{\delta \widehat A^a_\mu(x)} =
\G_{\Omega^a_\mu A^{*b}_\nu}(x,y) .
\label{def.1}
\eea
Once analyticity
in the background gauge field $\widehat A^a_\mu$ is assumed,
one can prove
that the dependence on the background field $\widehat A^a_\mu$ 
of the vertex functional {\it in the zero ghost sector} $\left . \G \right |_{c=0}$ 
is uniquely fixed by applying the 
transformation (\ref{bkg.split})
to $\left . \G \right |_{c=0}$ evaluated at $\widehat A_\mu=0$~\cite{Binosi:2011ar}.

The 1-PI amplitudes involving background insertions
can thus be obtained by those at zero background 
once the functional ${\cal G}^a_\mu$ is known. Notice, in fact, that the two-point function
$\G_{\Omega^a_\mu A^{*b}_\nu}$ can in principle be explored by means of non-perturbative methods, {\it e.g.}, on the lattice; in particular, in the Landau gauge it is related to the 1-PI connected part of a certain correlator which involves the time ordered product of two Faddeev-Popov determinants~\cite{Binosi:2011ar}.

From the physical point of view, besides the perhaps surprising fact that the background BRST invariance leads to such non-trivial consequences, these results entail the possibility of encoding topological information into continuum non-perturbative methods ({\it e.g.},  techniques based on the SD equations) through the systematic calculation of the correction terms due to the presence of a non-trivial background. In this way one might be able to describe what happens when topological effects are properly taken into accounts, and systematically study their effects on different correlators.

In this paper we generalize the  results of~\cite{Binosi:2011ar} to the ghost-dependent sector. This can be achieved in a natural and rather elegant way 
by means of a canonical transformation w.r.t. the Batalin-Vilkovisky (BV)
bracket associated with the ST identity. We will indeed show that, in the full quantum theory, the source $\Omega^a_\mu$ can be understood as the source coupled to the
generating functional of the canonical transformation that controls the (quantum-deformed) quantum-background splitting. With that will come the surprising feature that this canonical transformation also involves the ghost fields, contrary to the classical case in which the splitting is limited to the gauge sector.

Through the extension of the tools originally
devised for the direct imposition of the ST identity 
by algebraic methods \cite{at}, we will then devise the algebraic tools, required for obtaining an explicit, recursive representation
of the background-dependent sector of the vertex functional $\G$, based on homotopy techniques. This may prove useful in future practical computations, as it controls the corrections to the quantum $n$-point functions due to the presence of non-trivial backgrounds. Indeed, 
the canonical transformation gives rise to a field and antifield redefinition governed by
certain kernels involving the insertion of the $\Omega^a_\mu$ source, 
which can be computed non-perturbatively as solution of the corresponding SD equations.

We then make two examples of the possible use of the formalism.
To begin with, since the homotopy formula gives the explicit dependence of the vertex functional on $\widehat A^a_\mu$, we use it in order to derive the renormalization group (RG) equation in the presence of a non-trivial background. We then exploit this RG equation to obtain the value of the SU(2) Yang-Mills vertex functional on the instanton background at higher orders in the loop expansion, discussing in particular how the anomalous dimensions enter in the RG equation.

Finally, we compute the one-loop deformation of the quantum-background splitting in the case of a SU(2) instanton, in lowest order in the instanton background.

The paper is organized as follows.
In Section~\ref{sec.2} we set up our conventions
and write the extended ST identity by exploiting
the BV bracket. Next, we show that the extended ST
identity can be cast in the form of an inhomogeneous
equation for a suitable BRST differential acting on the
background field $\widehat A^a_\mu$ and its external ghost 
counterpart $\Omega^a_\mu$; in addition, we establish
the associated homotopy operator. Then, we construct the finite
canonical transformation which solves the
extended ST identity, fixing uniquely the dependence
on the background field in the $\Omega^a_\mu=0$ sector. 
In Section~\ref{sec.far} we analyze the canonical transformation
in terms of a field and antifield redefinition, controlled by certain
kernels involving the source $\Omega^a_\mu$. The SD equations
for these kernels are also given.
In Section~\ref{sec.rgi} we use the formalism to write down the RG equation in the presence of a generic background, discussing in particular the role of the anomalous dimensions of the gauge and the background fields.
We show how the formalism can be applied for an explicit background choice, corresponding to the celebrated BPST instanton~\cite{Belavin:1975fg}, in Section~\ref{sec.ex}. Specifically, we first evaluate the vertex functional on the instanton background by exploiting the RG equation previously derived; next,
we evaluate the one-loop corrections to the classical instanton profile.
Our conclusions are presented in Section~\ref{sec.concl}. The paper ends with an Appendix where we collect the tree-level vertex functional and the relevant functional identities of the theory.

\section{Canonical transformation for the quantum-background splitting}\label{sec.2}

\subsection{BV formulation of the ST identity}

We will adopt for the BV bracket the same conventions as in~\cite{Gomis:1994he}; then, using only left derivatives, one can write
\bea
(X,Y) = \int\!\diff^4x \sum_\phi
\left[ (-1)^{\epsilon_{\phi} (\epsilon_X+1)}
\frac{\delta X}{\delta \phi} \frac{\delta Y}{\delta \phi^*}
- (-1)^{\epsilon_{\phi^*} (\epsilon_X+1)}
\frac{\delta X}{\delta \phi^*} \frac{\delta Y}{\delta \phi}
\right],
\label{bracket}
\eea
where the sum runs over the fields $\phi = \{A^a_\mu,c^a \}$ and the antifields 
$\phi^* = \{ A^{*a}_\mu, c^{*a} \}$, and $\epsilon_\phi$, $\epsilon_{\phi^*}$ and $\epsilon_X$ represent  the statistics of the field $\phi$, the antifield $\phi^*$ and the functional $X$ respectively.
For convenience, a list of the ghost charge, statistics and mass dimension of the SU(N) Yang-Mills conventional fields and antifields together with the background fields and sources is given in Table~\ref{tableI}. The tree-level SU(N) vertex functional for an arbitrary background $R_\xi$ gauge is also given in Appendix~\ref{appendixA}.

\begin{table}
\begin{center}
\begin{tabular}{r||c|c|c|c|c|c||c|c|}
 & $\ A^a_\mu\ $ &  $\ c^a\ $ & $\ \bar c^a\ $  & $\ b^a\ $ &  $\ A^{*a}_\mu\ $ & $\ c^{*a}\ $  & $\ \widehat{A}^a_\mu\ $ & $\ \Omega^a_\mu\ $\\
\hline\hline
Ghost charge & 0  & 1 & -1  & 0  & -1  & -2 & 0 & 1\\
\hline
Statistics  & B & F & F  & B & F &  B & B & F\\
\hline
Dimension & 1 &  0 & 2 & 2 & 3 &  4  & 1 & 1 \\
\hline
\end{tabular} 
\caption{Ghost charge, statistics (B for Bose, F for Fermi), and mass dimension of both the SU(N) Yang-Mills conventional fields and antifields as well as background fields and sources. \label{tableI}}
\end{center}
\end{table}

Since the dependence on the Nakanishi-Lautrup field $b^a$ is confined at the classical level by the
$b$-equation~(\ref{b-equation}) one can  use 
 the reduced ($b$-independent) functional $\widetilde \G$ defined as
\bea
\widetilde \G = \G - \int\!\diff^4x \, b^a [\widehat {\cal D} (A-\hat A)]^a+\frac\xi2\int\!\diff^4x\,(b^a)^2.
\label{reduced}
\eea
At the same time, the fields  $b^a$ and $\bar c^a$ form a BRST doublet~\cite{Barnich:2000zw,Quadri:2002nh}, {\it i.e.}, a set of variables $u,v$
transforming under the BRST differential $s$ according to
 $su = v$, $sv=0$. This allows one to eliminate $\bar c^a$  through the redefinition $\widetilde A^{*a}_\mu = A^{*a}_{\mu} + (\widehat {\cal D}_\mu \bar c)^a$. Finally, since,
due to the antighost equation~(\ref{antighost-equation}), the vertex
functional depends on $\bar c^a$ only via the combination $\widetilde A^{*a}_\mu$, we will simply denote the latter combination by $A^{*a}_\mu $ in what follows. In the present paper we will always use the
reduced functional and hence we will just write $\G$
for $\widetilde \G$. For an alternative but equivalent formulation in which the fields $b^a$ and $\bar{c}^a$ are retained together with the corresponding antifields see~\cite{Quadri:2011aa}.

The extended ST identity in the presence of a
background field~\cite{Binosi:2011ar,Ferrari:2000yp} can then be written as
\be
\int\!\diff^4x\, \Omega^a_\mu(x)
\frac{\delta \G}{\delta \widehat A^a_\mu(x)} = 
- \frac{1}{2}\, (\G,\G) .
\label{m.1}
\ee
Notice that, in order to match the sign
conventions of Eq.~(\ref{bracket}), we have 
redefined $c^{*a} \rightarrow -c^{*a}$ as compared
with the choice of~\cite{Binosi:2011ar}.
By taking a derivative w.r.t. $\Omega^a_\mu$ and
then setting $\Omega^a_\mu=0$ we find 
%
\bea
\left.\frac{\delta \G}{\delta \widehat A^a_\mu(x)}\right|_{\Omega=0} = 
\left.- \left ( \frac{\delta \G}{\delta \Omega^a_\mu(x)},
\G \right)\right|_{\Omega=0} .
\label{m.2}
\eea

This is a very interesting equation.
It can be interpreted by saying that the 
derivative of the vertex functional w.r.t. 
the background field equals the 
effect of an infinitesimal canonical
transformation (w.r.t. the BV bracket)
on the vertex functional itself.
Notice that the BV bracket does depend
neither on $\widehat A^a_\mu$ nor on $\Omega^a_\mu$.
Thus, if we were able to write the finite
canonical transformation generated by
$\frac{\delta \G}{\delta \Omega^a_\mu}$, we would
control the full dependence of $\G$
on the background fields (also in the ghost-dependent
sector). We remark that Eq.~(\ref{m.2}) is valid not only for the counterterms  of $\G$ but for the full 1-PI Green's functions, and thus controls even the non-local dependence on the background.  

\subsection{Auxiliary BRST Differential and Homotopy Operator}

Although simple and natural from a
geometrical point of view,
the task of solving Eq.~(\ref{m.2}) is technically
rather involved and 
requires the extension of several algebraic
tools borrowed from cohomological
methods in gauge theories~\cite{Barnich:2000zw}.

For that purpose, it is convenient to introduce an auxiliary (nilpotent) BRST differential $\omega$ defined~as
\bea
\omega = \int\!\diff^4x\, \Omega^a_\mu(x)
\frac{\delta}{\delta \widehat A^a_\mu(x)};\qquad \omega^2=0.
\label{delta}
\eea
This differential maps $\widehat A^a_\mu$ into its BRST
partner $\Omega^a_\mu$, leaving all other fields and external sources unaltered.
The extended ST identity~(\ref{m.1}) can be then cast in the following form
\bea
\omega\, \G = - \frac{1}{2} (\G,\G);
\label{sti}
\eea
as a consequence of the nilpotency of $\omega$,
 one also finds the  consistency
condition  for the BV bracket of $\G$
\bea
\omega\, (\G,\G) = 0.
\label{s.1}
\eea

The advantage of this reformulation of the problem is that
one can use the homotopy operator $\kappa$, associated with $\omega$, in order to solve Eq.~(\ref{sti}). This operator is defined as~\cite{Zumino,Bettinelli:2007kc}
\bea
\kappa = \int_0^1\!\diff t \! \int\!\diff^4x \, \widehat A^a_\mu(x)
\lambda_t \frac{\delta}{\delta \Omega^a_\mu(x)},
\label{homotopy}
\eea
and fulfills the fundamental property
\bea
\{ \omega, \kappa \} = \left . \mathbb{I} \right |_{\Omega^a_\mu,\widehat A^a_\mu},
\label{id}
\eea
where the right-hand side (r.h.s.) represents the identity in the functional space spanned by monomials with at least one $\Omega^a_\mu$ or $\widehat A^a_\mu$. Finally, the operator $\lambda_t$ acts on a functional $X[\widehat A^a_\mu, \Omega^a_\mu; \zeta]$ (where $\zeta$ are fields and external sources
other than $\widehat A^a_\mu$ or $\Omega^a_\mu$) as
\bea
\lambda_t X[\widehat A^a_\mu,\Omega^a_\mu;\zeta] = 
X[t \widehat A^a_\mu, t \Omega^a_\mu; \zeta],
\label{lambda}
\eea
{\it i.e.}, it rescales by a factor $t$ the background 
field $\widehat A^a_\mu$ and its BRST partner
$\Omega^a_\mu$, leaving all other variables unchanged. 

One can easily write down a particular solution to Eq.~(\ref{sti}), and namely
\bea
\G = \G_0  - \frac{1}{2}\, \kappa\, (\G,\G),
\label{s.2.p}
\eea 
where $\G_0$ coincides with the vertex functional evaluated
at zero background field and therefore can be viewed as of setting the boundary condition for Eq.~(\ref{sti}).
Indeed, it is easy to show that~(\ref{s.2.p}) fulfills~(\ref{sti}), since, using
Eqs.~(\ref{s.1}) and~(\ref{id}), one has
\bea
\omega\, \G & = &-\frac{1}{2}\, \omega\kappa\, (\G,\G) =
-\frac{1}{2} \{ \omega, \kappa \} (\G,\G) +
\frac{1}{2} \,\kappa \omega\, (\G,\G)  =   -\frac{1}{2} (\G,\G).
\label{solution}
\eea

As usual, the most general solution of Eq.~(\ref{sti}) is obtained
by adding to the particular solution~(\ref{s.2.p}) the most general solution of the homogeneous equation
\bea
\omega X = 0.
\label{hom.sol}
\eea
Since $(\widehat A^a_\mu, \Omega^a_\mu)$ form
a BRST doublet, a general theorem
 in cohomology \cite{Barnich:2000zw,Quadri:2002nh} guarantees that 
 the most general solution to~(\ref{hom.sol}) is $\omega$-exact, {\it i.e.}, it must be generated by the $\omega$-variation of some
 functional $\Xi$: 
 \be
 X = \omega\, \Xi.
 \ee

Then, one can write the most general solution to Eq.~(\ref{sti}) in the following form
\be
\G = \G_0 + \omega\, \Xi - \frac{1}{2}\, \kappa\, (\G,\G) .
\label{s.2}
\ee
The ambiguities in the solution are controlled
by the $\omega$-exact term $\omega\, \Xi$; 
on the other hand, this term vanishes at $\Omega^a_\mu = 0$.
This is a very important point: the  
background-dependent amplitudes
that cannot be fixed uniquely by the ST identity~(\ref{sti}) do not affect the physically relevant
sector at $\Omega^a_\mu = 0$.
Hence in the latter sector we obtain the following representation
for the vertex functional:
\bea
\left . \G \right |_{\Omega= 0} = 
\left . \G_0 \right |_{\Omega = 0} - 
\frac{1}{2}\, \kappa \left . (\G, \G) \right |_{\Omega = 0}.
\label{s.3}
\eea
Since in what follows we will consider only the sector
$\Omega^a_\mu = 0$, we will refrain from writing explicitly that $\G$ should be calculated at $\Omega^a_\mu = 0$  whenever no confusion can arise.

Eq.(\ref{s.3}) is the basic homotopy formula allowing to control the dependence
of the vertex functional on the background field. Since it yields an explicit solution
to the extended ST identity~(\ref{m.1}), it is valid in any computational
framework in which the latter identity is fulfilled.
For instance it can be applied in the SD equations 
of non-perturbative QCD. 
Moreover it provides a strategy for the consistent implementation of the
background field method in lattice QCD, in the presence of a topologically
non-trivial background.

It should be emphasized that the homotopy formula separates the integration
over the quantum fluctuations of the gauge fields around the background
(accounted for by $\G_0$) from the background dependence of the
vertex functional, which can be recovered by purely algebraic means
through Eq.~(\ref{s.3}).

\subsection{\label{sec.3}Finite Canonical Transformation}

Using Eq.~(\ref{s.3}), we can derive a more explicit representation of the vertex functional $\G$. Substituting the explicit form~(\ref{homotopy}) of the operator $\kappa$, we get  
\be
\G = \G_0 - \int_0^1\!\diff t\!\int\!\diff^4x\, \widehat A^a_\mu(x)\, 
\lambda_t \left( \frac{\delta \G}{\delta \Omega^a_\mu(x)}, \G\right).
\label{s.5}
\ee 
Next, let us assume that $\G$ can be expanded in a power series in the background field  $\widehat A^a_\mu$ as 
\bea
\G = \sum_j \G_j,
\label{s.4}
\eea
with $\G_j$  the $j$th coefficient in the grading induced by the counting operator for $\widehat A^a_\mu$, or
\bea
{\cal N} \G_j = j \G_j ; \qquad
{\cal N} = \int\!\diff^4x \, \widehat A^a_\mu(x)
\frac{\delta}{\delta \widehat A^a_\mu(x)}.
\eea
One can then derive the first few coefficients of Eq.~(\ref{s.5})
in powers of $\widehat A^a_\mu$ as follows.
\begin{itemize}
\item At  zeroth order the r.h.s. of Eq.~(\ref{s.5}) reduces simply to
$\G_0$.

\item At first order we find
\bea
\G_1 & = & -\int_0^1\!\diff t \!\int\!\diff^4x\, \widehat A^a_\mu(x) \,
\lambda_t\, \left( \frac{\delta \G_0}{\delta \Omega^a_\mu(x)}, \G_0\right) \nonumber \\
& = & -\int\!\diff^4x \, \widehat A^a_\mu(x) 
 \left ( \frac{\delta \G_0}{\delta \Omega^a_\mu(x)}, \G_0\right).
\label{o.1}
\eea

\item At second order, two terms arise:
\bea
\G_2 & = & -\int_0^1\!\diff t \! \int\diff^4x \, \widehat A^a_\mu(x)\,
\lambda_t \left( \frac{\delta \G_1}{\delta \Omega^a_\mu(x)}, \G_0\right) \nonumber \\
& - &  \int_0^1\!\diff t\! \int\diff^4x \, \widehat A^a_\mu(x)\, 
\lambda_t \left( \frac{\delta \G_0}{\delta \Omega^a_\mu(x)}, \G_1\right) \nonumber \\
& = & 
-\frac{1}{2} \int\!\diff^4x \, \widehat A^a_\mu(x) 
 \left( \frac{\delta \G_1}{\delta \Omega^a_\mu(x)}, \G_0\right) - \frac{1}{2} \int\diff^4x \, \widehat A^a_\mu(x) 
 \left( \frac{\delta \G_0}{\delta \Omega^a_\mu(x)}, \G_1\right).
\label{o.2}
\eea
Inserting Eq.~(\ref{o.1})  in the second term of~(\ref{o.2}), we get
\bea
\G_2  & = &-\frac{1}{2} \int\!\diff^4x \, \widehat A^a_\mu(x) 
 \left( \frac{\delta \G_1}{\delta \Omega^a_\mu(x)}, \G_0\right) \nonumber \\
&&
+\frac{1}{2} \int\!\diff^4x\!\int\!\diff^4y \, \widehat A^a_\mu(x) 
\widehat A^b_\nu(y)
\left( \frac{\delta \G_0}{\delta \Omega^a_\mu(x)}, 
\left( \frac{\delta \G_0}{\delta \Omega^b_\nu(y)}, \G_0\right)\right).
\label{o.4}
\eea
\end{itemize}

Clearly, in Eq.~(\ref{o.4}) the second term fits to the (naively expected) pattern of an exponential, while the first one does not.  To understand where the obstruction to the exponentiation comes from, and, in passing, showing the advantages of the homotopy technique, it is useful to rederive Eq.~(\ref{o.4}) directly from the identity~(\ref{sti}).
For that purpose
we differentiate Eq.~(\ref{sti}) w.r.t. $\Omega^a_\mu$
to get
\bea
\frac{\delta \G}{\delta \widehat A^a_\mu(x)} = -
\Big ( \frac{\delta \G}{\delta \Omega^a_\mu(x)}, \G \Big ) 
+ \int\!\diff^4y\, \Omega^b_\nu(y)
\frac{\delta^2 \G}{\delta \Omega^a_\mu(x) 
\delta \widehat A^b_\nu(y)} \, .
\label{diff.1}
\eea
Next, we expand $\G$ at $\Omega^a_\mu=0$ as a power series around
$\widehat A^a_\mu=0$  ($\G$  is understood at $\Omega^a_\mu=0$):
\bea
\G[\widehat A] & = & \G[0] + 
\int\!\diff^4x\, 
\left . \frac{\delta \G}{\delta \widehat A^a_\mu(x)}
\right |_{\widehat A =0}\widehat A^a_\mu(x) \nonumber \\
&+&\frac12\int\!\diff^4x\!\int\!\diff^4y \, 
\left . \frac{\delta^2 \G}{\delta \widehat A^a_\mu(x)
\delta \widehat A^b_\nu(y)}
\right |_{\widehat A =0} 
\widehat A^a_\mu(x)
\widehat A^b_\nu(y) + \cdots
\label{diff.2}
\eea
The second term in the first line of the above equation
can be identified by setting $\widehat A^a_\mu=\Omega^a_\mu =0$ in Eq.~(\ref{diff.1})
\bea
\left . \frac{\delta \G}{\delta \widehat A^a_\mu(x)}
\right |_{\widehat A=0} = -
\left( \frac{\delta \G_0}{\delta \Omega^a_\mu(x)}, \G_0 \right).
\label{diff.3}
\eea
This result is in agreement with Eq.~(\ref{o.1}).

Then, let us differentiate Eq.~(\ref{diff.1}) w.r.t. $\widehat{A}^b_\nu(y)$ and set $\widehat A^a_\mu=\Omega^a_\mu =0$ afterwards. Since this derivative  can act inside the bracket on both
$\frac{\delta \G}{\delta \Omega^a_\mu}$ and $\G$, we obtain
\be
\left . 
\frac{\delta^2 \G}{\delta \widehat A^b_\nu(y)
\delta \widehat A^a_\mu(x)} 
\right |_{\widehat A= \Omega= 0} = 
- \left .\left ( \frac{\delta^2 \G}{\delta \widehat A^b_\nu(y) \Omega^a_\mu(x)}, \G \right ) \right |_{\widehat A = \Omega =0 } -
\left . 
\left ( \frac{\delta \G}{\delta \Omega^a_\mu(x)}, 
\frac{\delta \G}{\delta \widehat A^b_\nu(y)} \right) 
\right |_{\widehat A = \Omega = 0}.
\label{diff.4}
\ee
Upon substitution of~(\ref{diff.3}),
one obtains from the second term in the r.h.s.  the second term of Eq.~(\ref{o.4}), {\it i.e.}, the exponentiating one.
The first term is the non-exponentiating one; as can be clearly seen, it arises
from the dependence of the generating functional of the canonical transformation $\frac{\delta \G}{\delta \Omega^a_\mu}$ on the background field $\widehat A^a_\mu$. Thus it is this latter dependence 
that forbids to obtain a simple exponential as the solution to the finite canonical transformation (we will return on this point in Section~ref{sec.concl}).

Fortunately, however, the homotopy technique allows us
to write in a compact way all terms of the
finite canonical transformation associated with Eq.~(\ref{m.2}). 
This can be achieved by equating the $n$th coefficient (with $n>0$) in the background field of both sides of Eq.~(\ref{s.5}); one obtains then 
\be
\G_n = - \frac{1}{n} \int\!\diff^4x \, 
\widehat A^a_\mu(x) \sum_{k=0}^{n-1}
\left ( \frac{\delta \G_k}{\delta \Omega^a_\mu(x)},
\G_{n-1-k} \right).
\ee
This equation can be used iteratively in order to get
the terms $\G_n$ in the expansion of the
vertex functional.

\section{Field and antifield Redefinition and the SD equations}\label{sec.far}

In the previous sections we have focussed on finding a solution to the extended  ST identity~(\ref{sti}) recursively in the number of background fields. Here we will rather concentrate on seeing whether the solution can be generated by a suitable field and antifield redefinition which generalizes the classical background-quantum splitting. This has been already proven to be the case in the zero ghost sector~\cite{Binosi:2011ar}.

To this end, let us take a derivative w.r.t. $\Omega^a_\mu$ of Eq.~(\ref{sti}) and set $\Omega^a_\mu=0$ afterwards;
we get 
\bea
\frac{\delta \G}{\delta \widehat A^a_\mu} &=& - \int\!\diff^4x \left(
\left . \frac{\delta^2 \G}{\delta \Omega^a_\mu \delta A^{*b}_\nu} \right |_{\Omega=0} 
         \frac{\delta \G}{\delta A^b_\nu} -
\frac{\delta \G}{\delta A^{*b}_\nu} \left . \frac{\delta^2 \G}{\delta \Omega^a_\mu \delta A^b_\nu} \right |_{\Omega = 0}\right.
\nonumber \\
&-&\left . \frac{\delta^2 \G}{\delta \Omega^a_\mu\delta c^{*b}} \right |_{\Omega=0} \frac{\delta \G}{\delta c^b}-\left.
\frac{\delta \G}{\delta c^{*b}} \left . \frac{\delta^2 \G}{\delta \Omega^a_\mu \delta c^b} \right |_{\Omega = 0} 
\right) .
\label{redef.1}
\eea
Suppose now  that one can find a set of field and antifield redefinitions
\bea
A^a_\nu \rightarrow A^a_\nu - {\cal G}^a_\nu; &\quad& A^{*a}_\nu \rightarrow A^{*a}_\nu - {\cal G}^{*a}_\nu, 
\nonumber \\
c^a \rightarrow c^a + {\cal C}^a; &\qquad&  c^{*a} \rightarrow c^{*a} + {\cal C}^{a*} , 
\label{redef.2}
\eea
such that
\bea
\frac{\delta {\cal G}^b_\nu}{\delta \widehat A^a_\mu} = \left . \frac{\delta^2 \G}{\delta \Omega^a_\mu \delta A^{*b}_\nu} \right |_{\Omega=0}; &\qquad&
\frac{\delta {\cal G}^{*b}_\nu}{\delta \widehat A^a_\mu} = \left . \frac{\delta^2 \G}{\delta \Omega^a_\mu \delta A^{b}_\nu} \right |_{\Omega=0},  \nonumber \\
\frac{\delta {\cal C}^{b}}{\delta \widehat A^a_\mu} = \left . \frac{\delta^2 \G}{\delta \Omega^a_\mu \delta c^{*b}} \right |_{\Omega=0}; &\qquad& 
\frac{\delta {\cal C}^{*b}}{\delta \widehat A^a_\mu} = \left . \frac{\delta^2 \G}{\delta \Omega^a_\mu \delta  c^b} \right |_{\Omega=0}. 
\label{redef.3}
\eea
Then, the solution to Eq.~(\ref{redef.1}) is obtained by carrying out the field and antifield redefinition in Eq.~(\ref{redef.2}) on the vertex
functional at zero background $\G[A^a_\mu, c^a, A^{*a}_\mu, c^{*a};0]$ according to
\be
\G[A^a_\mu, c^a, A^{*a}_\mu, c^{*a}; \widehat A^a_\mu] = 
\G[A^a_\mu - {\cal G}^a_\mu, c^a + {\cal C}^a, A^{*a}_\mu - {\cal G}^{*a}_\mu, c^{*a} + {\cal C}^{*a}; 0].
 \ee
Taking a derivative of the  left-hand side w.r.t. $\widehat{A}^a_\mu$, and next using the chain rule on the r.h.s. while being careful about signs for fermionic variables, one can convince him/herself that the result would be precisely Eq.~(\ref{redef.1}) when the different resulting   terms are identified according to Eq.~(\ref{redef.3}).

The background-dependent field and antifield redefinition~(\ref{redef.2})
generalizes the classical background-quantum splitting and is the correct mapping
when quantum corrections are taken into
account. This result directly follows from the requirement of the validity of the ST identity.
We remark that the redefinition~(\ref{redef.2}) also involves the ghosts and the antifields. 
This is in sharp
contrast with the classical background-quantum splitting, which
is limited to the gauge field.

To lowest order in the background field,  Eqs.~(\ref{redef.2}) and~(\ref{redef.3}) give
\bea
A^a_\mu(x)&\to&A^a_\mu(x)-\int\!\diff^4y\,\Gamma_{\Omega^b_\nu A^{*a}_\mu}(y,x)\widehat{A}^b_\nu(y),\nonumber \\
c^a(x) &\to& c^a(x)+\int\!\diff^4y\int\!\diff^4z\,\G_{{\Omega^b_\mu}c^{*a}c^d}(y,x,z)\widehat{A}^{b}_\mu(y)c^d(z),\nonumber \\
A^{*a}_\mu(x)&\to&A^{*a}_\mu(x)-\int\!\diff^4y\int\!\diff^4z\,\Gamma_{A^{*d}_\rho\Omega^b_\nu A^{a}_\mu}(z,y,x){A}^{*d}_\rho(z)\widehat{A}^b_\nu(y),\nonumber \\
c^{*a}(x) &\to& c^{*a}(x)+\int\!\diff^4y\int\!\diff^4z\,\G_{{\Omega^b_\mu}c^ac^{*d}}(y,x,z)\widehat{A}^{b}_\mu(y)c^{*d}(z),
\label{low-or}
\eea
where the 1-PI functions are to be evaluated at $\widehat{A}^a_\mu=0$.
As can be seen, at this order there are only three independent functions that determine the splitting at the quantum level, since $c^{a}$ and $c^{*a}$ are controlled by one and the same function. In addition, from Table~\ref{tableI}, we see that the only superficially divergent term appears in the redefinition of $A^a_\mu$.

For the gauge field notice also that  the quantum background splitting $A^a_\mu=Q^a_\mu+\widehat{A}^a_\mu$ allows us to reinterpret the leading term in the field redefinition as a deformation of the background field, since
\be
Q^a_\mu(x)=A^a_\mu(x)-V^a_\mu(x);\qquad V^a_\mu(x)=\widehat{A}^a_\mu(x)+\int\!\diff^4y\,\Gamma_{\Omega^b_\nu A^{*a}_\mu}(y,x)\widehat{A}^b_\nu(y),
\ee
or, equivalently, in momentum space \be
V^a_\mu(p)=\left[g_{\mu\nu}\delta^{ab}+\Gamma_{\Omega^b_\nu A^{*a}_\mu}(p)\right]\widehat{A}^b_\nu(p).
\ee

\begin{figure}
\includegraphics[scale=.75]{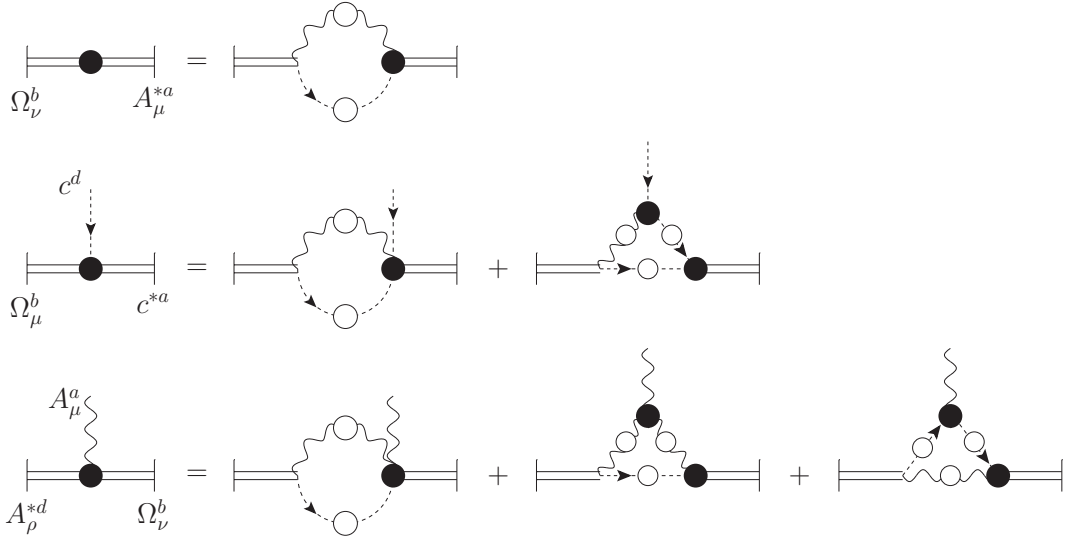}
\caption{\label{fig1}Schematic representation of the SD equations satisfied by the 1-PI functions $\Gamma_{\Omega^b_\nu A^{*a}_\mu}$, $\G_{\Omega^b_\mu c^{*a}c^d}$ and $\G_{A^{*d}_\rho\Omega^b_\nu A^a_\mu}$. White (black) blobs corresponds to connected (1-PI) Green's functions.}
\end{figure}

The SD equations that describe all the 1-PI functions appearing in the lowest order expansion~(\ref{low-or}) are shown in Fig.~\ref{fig1}. In particular, it should be noticed that the function $\Gamma_{\Omega^b_\nu A^{*a}_\mu}$ is the only one that has been studied in the literature~\cite{Grassi:2004yq}; in the Landau gauge it is related to the well-known Kugo-Ojima function~\cite{Kugo:1979gm}.

We conclude by observing that the existence of the field and antifield redefinitions of Eq.~(\ref{redef.2}) requires a careful check of the corresponding
integrability conditions. This has been already done for the case of the gauge field in~\cite{Binosi:2011ar} through an extensive use of the relations among 1-PI amplitudes encoded in the ST identity.
The analysis of the general case will be deferred to a later work;
here we remark that these redefinitions are related to the deformation of the canonical variables controlled by the canonical transformation generated by 
$\left . \frac{\delta\G}{\delta \Omega^a_\mu} \right |_{\Omega = 0}$.

\section{\label{sec.rgi}The renormalization group equation in the presence of non-trivial backgrounds}

The formalism developed so far imposes highly non-trivial constraints on the RG equation satisfied by the vertex functional $\Gamma$ in the presence of a background field 
$\widehat{A}^a_\mu \neq0$. 

To see this, let us start by considering a generic background $\widehat{A}^a_\mu$ depending on $N$ parameters $z_i$, with $i=1,\dots,N$. 
For instance, if $\widehat{A}^a_\mu$ is an
instanton background, $z_i$ are the instanton size, its
center and the coordinates of the relative orientation of
the instanton solution when $\widehat{A}^a_\mu$ is embedded
in a gauge group larger than SU(2).

We are interested in discussing the renormalization of the theory
at fixed background ({\it i.e.}, we do not perform the path-integral over the collective coordinates of the background, but only on the quantum fluctuations
of the gauge field).

On general grounds, the renormalization procedure might require
to renormalize  the $z_i$ parameters, which can be thought
of as additional couplings entering into the Feynman rules of the
model. To be sure, if this is the case and the parameters $z_i$ get renormalized, an additional dependence on the renormalization scale $\mu$ arises through the dependence of the $z_i$'s on $\mu$. Therefore,
the RG equation for the Yang-Mills theory
in a generic background $\widehat{A}^a_\mu$ takes the  form  
\bea
&& \Bigg [ 
\mu \frac{\partial}{\partial \mu} + 
\beta \frac{\partial}{\partial g}
+ \beta_i \frac{\partial}{\partial z_i}
+ \sum_{\phi \in \{ A, c \}}\gamma_\phi \int \diff^4x \,
\phi \frac{\delta}{\delta \phi} \nonumber \\
&&  \qquad +\
\gamma_{\widehat{A}} \int \diff^4x \, \widehat{A}^a_\mu \frac{\delta}{\delta A^a_\mu}
+ \gamma_{\widehat{A}}^{\mathrm{bkg}} \int \diff^4x \, 
\widehat{A}^a_{\mu} \frac{\delta}{\delta \widehat{A}^a_\mu} \Bigg] \G = 0 \, . 
\label{rg.1}
\eea
In the above equation $\beta$ is the usual $\beta$-function for the
coupling constant $g$, $\beta_i$ are the additional $\beta$-functions
for the parameters $z_i$, while $\phi$ is a collective notation for the ghost and
gauge fields with the corresponding anomalous dimensions $\gamma_\phi$.
$\gamma_{\widehat{A}}$ is the anomalous dimension associated with the
shift $A^a_\mu = \widehat{A}^a_\mu + Q^a_\mu$ and, finally, $\gamma_{\widehat{A}}^\mathrm{bkg}$ denotes
the anomalous dimension for the background field, which in general is also
possible and corresponds to a multiplicative renormalization of the
background, {\it i.e.}, $\widehat A^a_\mu \rightarrow Z^\mathrm{bkg} \widehat A^a_\mu$.

On the other hand, we know from our previous analysis that the
whole dependence on $\widehat{A}^a_\mu$ can be recovered by the canonical
transformation generated by $\frac{\delta \G}{\delta\Omega^a_\mu}$.
This fact is intimately related with the way used for introducing
the background field, {\it i.e.}, via a background gauge-fixing -- allowing us to write the gauge-fixing term as a BRST-exact
functional also in the presence of the background -- and the 
subsequent classical background-quantum splitting 
$A^a_\mu = \widehat{A}^a_\mu + Q^a_\mu$.

As already noticed, the
the generating functional of the canonical transformation
 $\frac{\delta \G}{\delta\Omega^a_\mu}$ contains the single divergent term $\G_{\Omega^a_\mu A^{*b}_\nu}$; therefore
all the dependence on the renormalization group scale $\mu$ in the
background-dependent sector will occur 
through this unique function. Moreover, one should also notice that 
this latter function need to be evaluated at zero external background, since the insertion of one or more background legs makes it superficially convergent.

These facts have two very important consequences. To begin with,
 the parameters $z_i$ characterizing the background
will not be subjected to any renormalization, {\it i.e.}, all the $\beta_i$ will be identically zero. Second, since the divergence of  $\G_{\Omega^a_\mu A^{*b}_\nu}$  at zero background is 
controlled by the single invariant
\bea
{\cal S}_0\left(\int\!\diff^4x\,  A^{a*}_{\mu} \widehat{A}^a_\mu\right) &=&
\int\!\diff^4x\, \left( \widehat{A}^a_\mu \frac{\delta \G^{(0)}}{\delta A^a_\mu} - A^{*a}_\mu \Omega^a_\mu\right),
\label{inv}
\eea
with ${\cal S}_0$ the usual linearized Slavnov-Taylor operator 
\be
{\cal S}_0=\int\!\diff^4x\,\left(\frac{\delta\Gamma^{(0)}}{\delta A^{*a}_\mu}\frac{\delta}{\delta A^a_{\mu}}+\frac{\delta\Gamma^{(0)}}{\delta A_{\mu}^a}\frac{\delta}{\delta A^{*a}_\mu}-\frac{\delta\Gamma^{(0)}}{\delta c^{*a}}\frac{\delta}{\delta c^{a}}-\frac{\delta\Gamma^{(0)}}{\delta c^a}\frac{\delta}{\delta c^{*a}}
+\Omega^a_\mu\frac{\delta}{\delta\widehat{A}^a_{\mu}}\right),
\ee
the RG equation will not display a term proportional to the background legs counting operator~$\widehat{A}^a_\mu\frac{\delta}{\delta\widehat{A}^a_\mu}$.

Thus for the Yang-Mills action in the gauge field sector one will get the result
\be
\left[\mu\frac{\partial}{\partial\mu}+\beta\frac{\partial}{\partial g}+\gamma_A\int\!\diff^4x\, A^a_\mu\frac{\delta}{\delta A^a_\mu}+\gamma_{\widehat{A}}\int\!\diff^4x\, \widehat{A}^a_\mu\frac{\delta}{\delta A^a_\mu}
\right]\Gamma=0.
\label{RGE}
\ee 

The important consequences of this RG equation will be analyzed in the following section in the particular case in which the background is chosen to be a SU(2) Yang-Mills instanton configuration.

\section{\label{sec.ex}An explicit example: the instanton background}

As a practical example of the many possible physical applications of the formalism developed, we consider the specific case in which the background is given by a single SU(2) Yang-Mills instanton. 

In order to establish the notation, let us indicate with $\widehat{A}^a_\mu$ the classical solution corresponding to the tree-level instanton profile in the singular gauge centered around the origin, which will be parametrized as (Euclidean space)
\be
\widehat{A}^a_\mu(x)=\overline{\eta}^a_{\mu\nu}x_\nu f_0(x); \qquad f_0(x)=\frac {2\rho^2}{x^2(x^2+\rho^2)},
\ee
where the (dimensionful) parameter  $\rho$ is the so-called instanton size.
Introducing then the (dimensionless) ratio $\lambda=r/\rho$ with $r=\sqrt{x_\mu x_\mu}$ the (tree-level) profile function can be rewritten as
\be
f_0(\lambda)=\frac 2{\rho^2}\frac1{\lambda^2(1+\lambda^2)}.
\label{treelev}
\ee
In momentum space, after defining
\be
\widehat{A}^a_\mu(p)=\overline{\eta}^a_{\mu\nu}p_\nu f_0(p); \qquad f_0(p)=-2\mathrm{i}\frac{\partial}{\partial p^2}\int\!\diff^4x\,\mathrm{e}^{\mathrm{i} p\cdot x}f_0(x),
\ee
we obtain, for the singular gauge instanton classical profile, the following expression
\bea
f_0(p)&=&\left(-8\pi^2\mathrm{i}\rho\right)\frac1{p^3}\left[-\frac2{p\rho}+K_1(p\rho)-(p\rho) K_1'(p\rho)\right]\nonumber \\
&=&\left(-8\pi^2\mathrm{i}\rho\right)\frac1{p^3}\left[-\frac2{p\rho}+(p\rho)K_2(p\rho)\right],
\label{effe}
\eea
with $K_i$ the modified Bessel functions of the second kind.

\subsection{Renormalization group analysis}

As is clear from the previous subsection, for a single instanton background 
centered around the origin there is only one dimensionful parameter, that 
is the instanton size $\rho$. Thus one can trade the dimensionful 
parameter $\mu$ for the dimensionless parameter $\zeta=\mu\rho$, so that 
the RG equation\noeq{RGE} will read in this case
\be
\left[\zeta\frac{\partial}{\partial \zeta}+\beta\frac{\partial}{\partial g}+\gamma_A\int\!\diff^4x\, A^a_\mu\frac{\delta}{\delta A^a_\mu}+\gamma_{\widehat{A}}\int\!\diff^4x\, \widehat{A}^a_\mu\frac{\delta}{\delta A^a_\mu}
\right]\Gamma=0.
\ee
In the one-loop approximation, one obtains
\be
\zeta\frac{\partial\Gamma^{(1)}}{\partial \zeta}+\beta_1\frac{\partial\Gamma^{(0)}}{\partial g}+\gamma^{(1)}_{A}\int\!\diff^4x\, A^a_\mu\frac{\delta\Gamma^{(0)}}{\delta A^a_\mu}+\gamma^{(1)}_{\widehat{A}}\int\!\diff^4x\, \widehat{A}^a_\mu\frac{\delta\Gamma^{(0)}}{\delta A^a_\mu}=0.
\label{RGE-1}
\ee

We would like to use this equation to evaluate the one loop vertex functional on the instanton configuration; to that end, we set $A^a_\mu=\widehat{A}^a_\mu$ in~\eq{RGE-1} while setting to zero all other fields and sources.  Taking then into account that the instanton is a solution of the classical Yang-Mills equation of motion, we find the final result
\be
\left. \zeta\frac{\partial\Gamma^{(1)}}{\partial \zeta}\right\vert_{A=\widehat{A}}+\left.\beta^{(1)}\frac{\partial\Gamma^{(0)}}{\partial g}\right\vert_{A=\widehat{A}}=0,
\ee
or, setting $\tau=\log \zeta$ and using the value of the tree-level action evaluated on the instanton configuration, 
\be
\left.\frac{\partial\Gamma^{(1)}}{\partial \tau}\right\vert_{A=\widehat{A}}=-\beta^{(1)}\frac{\partial}{\partial g}\frac{8\pi^2}{g^2}.
\ee
The solution is 
\be
\Gamma^{(1)}(\tau)=\Gamma^{(0)}(g+g^{(1)})+d^{(1)} +{\cal O}(\hbar^2),
\ee
with $d^{(1)}$ a $\tau$-independent constant that can be reabsorbed into a finite one-loop renormalization of the coupling $g$. 
$g^{(1)}$ is the one-loop coefficient of the renormalized coupling
constant, obeying
\bea
\frac{\partial g^{(1)}}{\partial \tau} = - \beta^{(1)} \, .
\label{beta.1}
\eea
This result therefore states nothing but the classic result of 't Hooft~\cite{'tHooft:1976fv}, {\it i.e.}, that the one-loop effects of the quantum corrections around the instanton profile resum in such a way that the net effect is the appearance of the one-loop $\beta$ function.  

At the two-loop level one has instead
\bea
&& \left.\zeta\frac{\partial\Gamma^{(2)}}{\partial \zeta}\right\vert_{A=\widehat{A}}+\left.\beta^{(2)}\frac{\partial\Gamma^{(0)}}{\partial g}\right\vert_{A=\widehat{A}}+\left.\beta^{(1)}\frac{\partial\Gamma^{(1)}}{\partial g}\right\vert_{A=\widehat{A}}
\nonumber \\
&& \qquad +\left.\gamma^{(1)}_{A}\int\!\diff^4x\, \widehat{A}^a_\mu\frac{\delta\Gamma^{(1)}}{\delta A^a_\mu}\right\vert_{A=\widehat{A}}
+\left.\gamma^{(1)}_{\widehat{A}}\int\!\diff^4x\, \widehat{A}^a_\mu\frac{\delta\Gamma^{(1)}}{\delta A^a_\mu}\right\vert_{A=\widehat{A}}=0,
\label{rg.2loop}
\eea
and we clearly see that the obstruction for a direct generalization of the one-loop result resides in the terms in the second line of the above equation, since the anomalous dimensions $\gamma_A, \gamma_{\widehat A}$ are
 in general non-vanishing.

The fact that RG-invariance of the ratio 
$R = \frac{\langle 0 | 0 \rangle_I}{\langle 0 | 0 \rangle}$ 
of the vacuum-to-vacuum amplitude in the
presence of an instanton over the vacuum-to-vacuum amplitude at
zero background does not hold at the two-loop level in the
single instanton approximation, as a consequence of the
anomalous dimension terms in Eq.~(\ref{rg.2loop}), has
been already noticed long  ago~\cite{Morris:1984zi} through explicit diagrammatic computations. In those papers it was found that the $\mu$-dependence
of $R$ in the single instanton approximation is canceled out
once the path-integral over the collective coordinates is carried out
with the appropriate extended Feynman rules, involving the ghosts
associated with the zero modes of the two-point gauge function
in the presence of the instanton. 

The advantage of the analysis presented here is that it has a simple and direct generalization to all orders; in addition, it can be 
combined with the algebraic treatment of the gauge field zero modes to obtain the appropriate RG equation when the collective coordinates are promoted to quantized fields~\cite{Amati:1978wu}.

\subsection{One-loop deformation of the instanton profile}

As a second example, we calculate the one-loop corrections to the instanton profile function. To the best of our knowledge this is the first time that such deformation is computed.

The function $\Gamma_{\Omega^a_\mu A^{*b}_\nu}$ (Fig.~\ref{fig1} first row) reads~\cite{Binosi:2007pi}
\be
\Gamma_{\Omega^b_\nu A^{*a}_\mu}(p)=gf^{bmn}g_{\nu\rho}\int_k\!D^{mm'}(k+p)\Delta_{nn'}^{\rho\rho'}(k)\Gamma_{c^{m'}A^{n'}_{\rho'}A^{*a}_{\mu}}(-k,-p);\qquad \int_k\equiv\mu^\epsilon\int\!\frac{\diff^dk}{(2\pi)^d},
\ee
with $\epsilon=4-d$ and $d$ the space-time dimension, while $\Delta$ and $D$ represents the all-order gluon and ghost propagators respectively. 
Introducing then the Lorentz and color decomposition
\be
\Gamma_{\Omega^b_\nu A^{*a}_\mu}(p)=-\frac{g^2C_A}{16\pi^2}\delta^{ab}\left[A(p)g_{\mu\nu}+B(p)\frac{p_\mu p_\nu}{p^2}\right],
\ee
we see that in the instanton case the $B$ form factor does not contribute. 

At the one-loop level we then obtain the
deformed background field
\be
V^a_\mu(p)=\overline{\eta}^a_{\mu\nu}p_\nu \left[f_0(p)+f_1(p)\right];\qquad f_1(p)=-\frac{g^2C_A}{16\pi^2}A^{(1)}(p)f_0(p),
\label{1l-mom}
\ee
where $f_0$ is given in Eq.~(\ref{effe}).\\
In the Landau gauge (which is the appropriate choice in the instanton case) one has
\be
\Gamma^{(1)}_{\Omega^b_\nu A^{*a}_\mu}(p)=-g^2C_A\delta^{ab}\int_k\frac1{k^2(k+p)^2}P_{\mu\nu}(k),
\ee
where $C_A$ is the Casimir eigenvalue of the adjoint representation
($C_A=N$ for SU(N)); a straightforward calculation gives (Euclidean space)
\bea
A^{(1)}(p)&=&-\frac32\frac1{d-4}+\frac32-\frac34\log\left(\frac{p^2}{\mu^2}\right),
\nonumber\\
B^{(1)}(p)&=&-\frac12.
\label{g.astar.omega}
\eea
Notice that  in the one-loop approximation this result is not affected by the inclusion of fermions in the theory.
The first term appearing in $A^{(1)}$ is clearly divergent in the $d\to4$ limit; this divergence is controlled by the invariant shown in \eq{inv} and therefore can be safely absorbed in the corresponding counterterm.

By evaluating the inverse Fourier transform of $f_1$ one obtains the quantum corrected instanton profile in position space:
\be
V^a_\mu(x)=\overline{\eta}^a_{\mu\nu}x_\nu\left[f_0(x) + f_1(x)\right];\qquad f_1(x)=\frac{\mathrm{i}}{4\pi^2}\frac{x_\nu}{r^2}\frac{\partial}{\partial x_\nu}\int_0^\infty\!\diff p\,p^3f_1(p)\frac1{pr}J_1(pr).
\ee
The evaluation of $f_1(x)$ can be performed analytically, and we find\footnote{This is only true in the singular gauge. In the regular gauge the integral over $p$ does not converge.}
\bea
f_1(x)&=&-3\frac{g^2C_A}{16\pi^2}\left[\frac1{\rho^2}\frac{1+\log\rho\mu}{\lambda^2(1+\lambda^2)}
-\frac{x_\nu}{r^2}\frac{\partial}{\partial x_\nu}\int_0^\infty\!\diff t\,  F(t,\lambda)\right], 
\eea
where we have  set $t=p\rho$ and
\be
F(t,\lambda)=\log t\left[-\frac2t+tK_2(t)\right]\frac1{\lambda t}J_1(\lambda t).
\ee
The integral in $t$ yields
\bea
\int_0^\infty\!\diff t\,  F(t,\lambda) &=& \frac{1}{8\lambda^2}
\left\{ \log^2(1+\lambda^2) \lambda^2 -
4 \left( \log \frac{\lambda}{4} +2 \gammaE - 1 \right)
\lambda^2 \log \lambda 
+2 \lambda^2  {\rm Li}_2 \left( \frac{1}{1+\lambda^2} \right)\right.\nonumber\\
&+&\left. \left[  -2\lambda^2 \log \frac{\lambda^2}{1+\lambda^2}  +
\left(-2 + 4 \gammaE - 4 \log 2\right) \lambda^2 - 2\right] 
\log \left(1+\lambda^2\right) 
\right\},
\label{math}
\eea
where $\gammaE$ is the Euler-Mascheroni constant ($\gammaE=0.57721\dots$) and ${\rm Li}_2$ is the standard dilogarithm. Thus one has
\be
\frac{x_\nu}{r^2}\frac{\partial}{\partial x_\nu}\int_0^\infty\!\diff t\,  F(t,\lambda) =\frac1{\rho^2}\left[-\frac{\gammaE-\log2}{\lambda^2(1+\lambda^2)}-\frac{\log\lambda}{\lambda^2}+\frac{1+\lambda^4}{2\lambda^4(1+\lambda^2)}\log(1+\lambda^2)\right],
\ee
which finally gives for $f_1$
\bea
f_1(\lambda)&=&-3\frac{g^2C_A}{16\pi^2}\frac1{\rho^2}\left[\frac{1+\log\rho\mu}{\lambda^2(1+\lambda^2)}+\frac{\gammaE-\log2}{\lambda^2(1+\lambda^2)}+\frac{\log\lambda}{\lambda^2}-\frac{1+\lambda^4}{2\lambda^4(1+\lambda^2)}\log(1+\lambda^2)\right]\!.\hspace{.8cm}
\eea

There are a number of comments that one can make regarding this result, and namely:
\begin{itemize}

\item Clearly the one-loop corrected instanton is neither self-dual nor it reduces to pure gauge as $r\to\infty$;

\item With the generic parametrization $V^a_\mu(x)=\overline{\eta}^a_{\mu\nu}x_\nu f(r)$ the field strength becomes
\be
F^a_{\mu\nu}=\overline{\eta}^a_{\mu\nu}\left [r^2 f^2(r) - 2 f(r)\right]- \left(\overline{\eta}^a_{\mu \rho} x_\nu x_\rho - \overline{\eta}^a_{\nu\rho} x_\mu x_\rho\right)
\left[\frac{f'(r)}{r} + f^2(r) \right],
\ee
which gives for the (Euclidean) Yang-Mills action 
\bea
S_{\s{\mathrm{Y\!M}}}&=&\frac14\int\!\diff^4x\,(F^a_{\mu\nu})^2\nonumber \\
&=&
2\pi^2\int_0^\infty\diff{r}\,r^3\left[\frac32 r^4 f^4(r) -6 r^2 f^3(r) +12 f^2(r) +6 r f(r) f'(r) +\frac32 r^2  f'(r)^2\right].\nonumber \\
\label{SYM}
\eea
When the r.h.s. of the above equation is expanded according to the loop
order, we see that our correction resums a particular subset of diagrams 
which are bound to contribute up to four loops;

\item For small $r$ the Yang-Mills action density in~(\ref{SYM}) calculated on the corrected profile goes like $1/r^4$ times logs; once multiplied by the $r^3$ coming from the measure, this leaves us with a log squared singularity for $r\sim0$ (that is either when $r\to0$ or $\rho\to\infty$). This is the usual infrared disease of instanton calculus that would be effectively cured by the dynamical generation of a gluon mass~\cite{Cornwall:1981zr}, firmly established recently in both lattice simulation~\cite{Cucchieri:2007md} as well as  SD studies of the gluon propagator $\Delta$~\cite{Aguilar:2008xm}. This would furnish a cutoff for the $r$ integral of the order $\Delta^{-1}(0)$;

\item Finally, it is interesting to notice that with respect to the tree-level profile, $f_1$ shows a log enhancement in both the small ($\lambda\to\infty$) and large ($\lambda\to0$) size limit. Due to these enhancements it is  tempting to conjecture that 
the contribution to $S_{\s{\mathrm{Y\!M}}}$ coming from the quantum-corrected instanton is larger than its classical counterpart
in both the infrared and ultraviolet regime.
But then the factor ${\mathrm e}^{-S_{\s{\mathrm{Y\!M}}}}$ would be smaller
for small as well as large size instantons, giving rise to a suppression for the instanton density in these two regimes. Though this is precisely what is observed on the lattice~\cite{Michael:1995br}, we remark that the large size limit lies beyond the validity of our perturbative result for~$f_1$.  
\end{itemize}

\section{\label{sec.concl}Discussion and Conclusions}

In this paper we have shown that the full dependence
of the vertex functional $\G$ on the background field
$\widehat{A}^a_\mu$ can be recovered by an appropriate
field redefinition generated by a canonical transformation
w.r.t. the BV bracket naturally associated with the
ST identity of the theory.
The BRST partner $\Omega^a_\mu$ of the background
field $\widehat{A}^a_\mu$ has been identified as the source coupled
to the fermionic generator of the infinitesimal canonical transformation; in addition, we were able to provide a recursive
formula for solving the finite canonical transformation
by making use of homotopy techniques.

As for the failure of the exponentiation of the solution for the finite canonical transformation (which has been ultimately traced back to the dependence of the generating functional $\frac{\delta\Gamma}{\delta\Omega^a_\mu}$ on the background field $\widehat{A}^a_\mu$), we notice that there is an analogy in classical mechanics. Indeed, suppose we want to describe the time evolution of some
function $f$, governed by the equation
\be
\frac{df}{dt} = \{ f, H \},
\label{cl.1}
\ee
where $\{\cdot , \cdot\}$ is the Poisson bracket.
Then, if $H$ is time-independent, the finite canonical
transformation generated by $H$ can be written
as an exponential
\bea
f(t) =f \exp(\widehat H t) \vert_0,
\eea
where $\widehat H$ is the operator $\{\cdot,H\}$, and the zero
denotes that all terms in the series on the r.h.s.
have to be evaluated at $t=0$. If, on the other hand, $H$ is time-dependent, further terms
in general arise and the finite canonical transformation
is more complicated. A general technique for constructing the mapping between the new and the old canonical variables when the generator depends on one parameter is known~\cite{Deprit:1969}; one might then ask if this approach can be extended to our case and thus used to obtain an explicit form of the field and antifield redefinitions of Eqs.~(\ref{redef.2}) and~(\ref{redef.3}). 

We have also shown how these formal techniques can be proficiently applied in practical physical situations.
In particular, we have derived the generic form of the RG equation in the presence of a background field. Once specializing to the case of a SU(2) Yang-Mills instanton, the classic one-loop result of 't Hooft is recovered; at the two-loop level, our equation allows for the systematic disentanglement of the contribution due to the fields anomalous dimensions which have been discussed in the literature only on a diagrammatic basis. Our approach could also be directly  extended  to all orders and applied in a situation where one performs the path integral over the quantized collective modes through the addition of the appropriate ghost fields~\cite{Amati:1978wu}.
Finally, in the single instanton approximation, we were able to  determine analytically the lowest order correction to the instanton profile both in momentum as well as in position space. Once inserted in the Yang-Mills action this amounts to take into account the resummation effects of a particular set of diagrams up to four loops.

\acknowledgements

It is a pleasure to thank E. Shuryak and A. Slavnov for useful discussions.

\appendix

\section{\label{appendixA}Tree-level vertex functional}

The tree-level vertex functional is written as
\bea
\G^{(0)}&=&\int\!\mathrm{d}^4x\left[-\frac14F^a_{\mu\nu} F^{a\mu\nu}-\bar{c}^a(\widehat{\cal D}_\mu{\cal D}^\mu c)^a-({\cal D}^\mu \bar{c})^a\Omega^a_\mu-\frac\xi2(b^a)^2+b^a[{\cal D}^\mu(A-\widehat{A})_\mu]^a\right.\nonumber \\
&+&\left.A^{*a}_\mu\left({\cal D}^\mu c\right)^a+\frac12f^{abc} c^{*a}c^bc^c\right],
\label{tlvf}
\eea
where the covariant derivative ${\cal D}$ is defined according to
\be
({\cal D}_\mu\phi)^a={\cal D}^{ab}_\mu\phi^b;\qquad 
{\cal D}^{ab}_\mu=\delta^{ab}\partial _\mu +f^{acb}A^c_\mu
\ee
($\widehat{\cal D}$ can be obtained from the above substituting $A^c_\mu$ with $\widehat{A}^c_\mu$).

The $b$-equation at the level of the {\it complete} vertex functional $\G$ reads
\be
\frac{\delta\Gamma}{\delta b^a}=-\xi b^a+[{\cal D}^\mu(A-\widehat{A})_\mu]^a,
\label{b-equation}
\ee
while the antighost equation is given by
\be
\frac{\delta\Gamma}{\delta {\bar{c}^a}}=-\widehat{\cal D}^{ab}_\mu\frac{\delta\Gamma}{\delta {A^{*b}_\mu}}+({\cal D}^\mu \Omega_\mu)^a.
\label{antighost-equation}
\ee

Finally, the Ward identity that holds in the background
gauge as a consequence of the invariance under background
gauge transformations reads
\be
{\cal W}^a(\G) = -\widehat{\cal D}^{ab}_\mu\frac{\delta \G}{\delta \widehat{A}^b_\mu}-\sum_{\chi}f^{abc} \chi^b \frac{\delta \fg}{\delta \chi^c} = 0,
\label{wti}
\ee
where $\chi$ runs over the fields $Q^a_\mu=A^a_\mu-\widehat{A}^a_\mu$, $c^a,\bar{c}^a,b^a$, the source $\Omega^a_\mu$ and the antifields $A^{*a}_\mu$ and $c^{*a}$.


\begin{thebibliography}{99}

\bibitem{DeWitt:1967ub}
  B.~S.~DeWitt,
  Phys.\ Rev.\  {\bf 162}, 1195  (1967);
  J.~Honerkamp,
  Nucl.\ Phys.\  B {\bf 48}, 269  (1972);
  R.~E.~Kallosh,
  Nucl.\ Phys.\  B\ {\bf 78}, 293  (1974);
  H.~Kluberg-Stern, J.~B.~Zuber,
  Phys.\ Rev.\ D\  {\bf 12}, 482  (1975);
  I.~Y.~.Arefeva, L.~D.~Faddeev, A.~A.~Slavnov,
  Theor.\ Math.\ Phys.\  {\bf 21}, 1165 (1975);
G.~'t~Hooft, {The Background Field Method in Gauge Field Theories, }In *Karpacz
  1975, Proceedings, Acta Universitatis Wratislaviensis No.368, Vol.1*, Wroclaw
 345 (1976);
  S.~Weinberg,
  Phys.\ Lett.\  B\ {\bf 91}, 51  (1980);
  G.~M.~Shore,
  Annals Phys.\  {\bf 137}, 262  (1981);
  L.~F.~Abbott, M.~T.~Grisaru, R.~K.~Schaefer,
  Nucl.\ Phys.\  B\ {\bf 229}, 372  (1983);
  C.~F.~Hart,
  Phys.\ Rev.\ D\ {\bf 28}, 1993 (1983).

\bibitem{Abbott:1980hw}
L.~F.~Abbott,
  Nucl.\ Phys.\  B {\bf 185}, 189 (1981);
  Acta Phys.\ Polon.\  B\ {\bf 13}, 33 (1982).

\bibitem{Ichinose:1981uw}
  S.~Ichinose, M.~Omote,
  Nucl.\ Phys.\ B\  {\bf 203}, 221 (1982);
  D.~M.~Capper, A.~MacLean,
  Nucl.\ Phys.\ B\  {\bf 203}, 413  (1982).

\bibitem{Denner:1994xt}
  A.~Denner, G.~Weiglein, S.~Dittmaier,
  Nucl.\ Phys.\ B \ {\bf 440}, 95  (1995).

\bibitem{hep-ph/0102005}
  P.~A.~Grassi, T.~Hurth and M.~Steinhauser,
  Nucl.\ Phys.\ B\ {\bf 610}, 215 (2001).
\bibitem{Gates:1983nr}
  S.~J.~Gates, M.~T.~Grisaru, M.~Rocek {\it et al.},
  Front.\ Phys.\  {\bf 58}, 1  (1983).

\bibitem{Ferrari:2000yp}
  R.~Ferrari, M.~Picariello, A.~Quadri,
  Annals Phys.\  {\bf 294 },  165-181 (2001).

\bibitem{Belavin:1975fg}
  A.~A.~Belavin, A.~M.~Polyakov, A.~S.~Schwartz, Y.~S.~Tyupkin,
  Phys.\ Lett.\  {\bf B59}, 85  (1975).

\bibitem{'tHooft:1976fv}
  G.~'t Hooft,
  Phys.\ Rev.\  {\bf D14}, 3432-3450 (1976).

\bibitem{dual_superconductor}
  Y.~Nambu,
  Phys.\ Rev.\ D\ {\bf 10}, 4262  (1974);
  G.~Parisi,
  Phys.\ Rev.\ D\ {\bf 11}, 970  (1975);
  S.~Mandelstam,
  Phys.\ Lett.\ B\ {\bf 53}, 476  (1975);
  A.~M.~Polyakov,
  Nucl.\ Phys.\ B\ {\bf 120}, 429  (1977);
  G.~'t Hooft,
  Nucl.\ Phys.\ B\ {\bf 153}, 141 (1979);
  G.~'t Hooft,
  Nucl.\ Phys.\ B\ {\bf 190}, 455  (1981).

\bibitem{vortex_condensation}
  G.~'t Hooft,
  Nucl.\ Phys.\ B\ {\bf 138}, 1 (1978);
  Y.~Aharonov, A.~Casher and S.~Yankielowicz,
  Nucl.\ Phys.\ B\ {\bf 146}, 256  (1978);
  J.~M.~Cornwall,
  Nucl.\ Phys.\ B\ {\bf 157}, 392(1979);
  H.~B.~Nielsen and P.~Olesen,
  Nucl.\ Phys.\ B\ {\bf 160}, 380 (1979).

\bibitem{Cornwall:1981zr}
J.~M.~Cornwall,
Phys.\ Rev.\ D {\bf 26}, 1453  (1982);
  J.~M.~Cornwall and J.~Papavassiliou,
  Phys.\ Rev.\  D\ {\bf 40}  (1989) 3474.

\bibitem{Binosi:2002ft}
  D.~Binosi and J.~Papavassiliou,
  Phys.\ Rev.\ D {\bf 66}(R), 111901 (2002);
  J.\ Phys.\ G {\bf 30}, 203   (2004); 
for a recent review on the subject see also 
  Phys.\ Rept.\  {\bf 479}, 1  (2009).

\bibitem{Binosi:2007pi}
  D.~Binosi and J.~Papavassiliou,
  Phys.\ Rev.\  D {\bf 77}(R), 061702 (2008);
  JHEP\ {\bf 0811}, 063   (2008).

\bibitem{Aguilar:2008xm}
A.~C.~Aguilar, D.~Binosi and J.~Papavassiliou,
Phys.\ Rev.\  D {\bf 78}, 025010 (2008).

\bibitem{Binosi:2011ar}
  D.~Binosi, A.~Quadri,
  Phys.\ Rev.\ D {\bf 84}, 065017 (2011)

\bibitem{Grassi:1995wr}
  P.~A.~Grassi,
  Nucl.\ Phys.\  B\ {\bf 462}, 524 (1996).

\bibitem{Becchi:1999ir}
  C.~Becchi, R.~Collina,
  Nucl.\ Phys.\  B\ {\bf 562}, 412  (1999).

\bibitem{at}
  A.~Quadri,
  JHEP\ {\bf 0304}, 017 (2003);
  J.\ Phys.\ G\ {\bf 30}, 677 (2004);
  JHEP\ {\bf 0506}, 068 (2005).

\bibitem{Gomis:1994he}
  J.~Gomis, J.~Paris, S.~Samuel,
  Phys.\ Rept.\  {\bf 259 }, 1-145 (1995).

\bibitem{Quadri:2011aa} 
  A.~Quadri,
  ``Background field dependence from the Slavnov-Taylor identity in (non-perturbative) Yang-Mills theory,'' {\it Proceedings of the  International Workshop on QCD Green's Functions, Confinement and Phenomenology}, Trento 5-9 September 2011 (PoS to appear), arXiv:1112.1817 [hep-th].

\bibitem{Barnich:2000zw}
  For a review see, {\it e.g.}, G.~Barnich, F.~Brandt, M.~Henneaux,
  Phys.\ Rept.\  {\bf 338 },   439-569 (2000).

\bibitem{Quadri:2002nh}
  A.~Quadri,
  JHEP {\bf 0205 } 051 (2002).
    
\bibitem{Zumino}
  B.~Zumino, Lectures given at Les Houches Summer School
  on Theoretical Physics, Les Houches, France, Aug 8 - Sep 2, 1983.

\bibitem{Bettinelli:2007kc}
  D.~Bettinelli, R.~Ferrari, A.~Quadri,
  JHEP {\bf 0703 } (2007)  065.

\bibitem{Grassi:2004yq} 
  P.~A.~Grassi, T.~Hurth and A.~Quadri,
  Phys.\ Rev.\ D {\bf 70}, 105014 (2004);
  A.~C.~Aguilar, D.~Binosi and J.~Papavassiliou,
  JHEP {\bf 0911}, 066 (2009), and references therein

\bibitem{Kugo:1979gm} 
  T.~Kugo and I.~Ojima,
  Prog.\ Theor.\ Phys.\ Suppl.\  {\bf 66}, 1 (1979).

\bibitem{Morris:1984zi}
  T.~R.~Morris, D.~A.~Ross, C.~T.~Sachrajda,
  Nucl.\ Phys.\  {\bf B255 }, 115 (1985);
  Phys.\ Lett.\  {\bf B158 }, 223 (1985);
  Nucl.\ Phys.\  {\bf B264 }, 111  (1986);
  Phys.\ Lett.\  {\bf B172 }, 40 (1986).

\bibitem{Amati:1978wu}
  D.~Amati, A.~Rouet,
  Nuovo Cim.\  {\bf A50 }, 265 (1979).

\bibitem{Cucchieri:2007md}
  A.~Cucchieri, T.~Mendes,
  PoS {\bf LAT2007}, 297 (2007);
for a recent review see also A.~Cucchieri, T.~Mendes,
  PoS {\bf QCD-TNT09}, 026 (2009).
 
\bibitem{Michael:1995br} 
  See, {\it e.g.}, C.~Michael and P.~S.~Spencer,
  Phys.\ Rev.\ D {\bf 52}, 4691 (1995).

\bibitem{Deprit:1969}
 A. Deprit,
 Celestial Mechanics and Dynamical Astronomy {\bf 1}, 12 (1969). 

\end{thebibliography}
\end{document}